\documentclass[namedreferences]{SolarPhysics}
\usepackage[optionalrh]{spr-sola-addons} 
\usepackage{epsfig}          
\usepackage{graphicx}        
\usepackage{color}           
\usepackage{url}             

\newcommand{\spn}{\,--\,}  
\newcommand{\sph}{-}         

\newcommand{\deriv}[2]{\frac{{\mathrm d} #1}{{\mathrm d} #2}}

\def\tm{\leavevmode\hbox{$\rm {}^{TM}$}}

\newcommand{\etal}{{\it et al.}}
\newcommand{\ie}{{\it i.e.}}
\newcommand{\cf}{{\it cf.}}
\newcommand{\eg}{{\it e.g.}}
\newcommand{\etc}{{\it etc}}

\begin{document}
\begin{article}
\begin{opening}

%
%

\title{A Technique for Automated Determination of Flare-ribbon Separation and Energy Release}

\author{R.A.~\surname{Maurya}$^{1}$\sep
        A.~\surname{Ambastha}$^{2}$
         }
\runningauthor{R.A. Maurya and A. Ambastha}
\runningtitle{Flare-ribbon Separation and Energy Release}

\institute{$^{1}$Udaipur Solar Observatory (Physical Research Laboratory),\\ P.O. Box 198, Dewali, Badi Road, Udaipur 313 001, India.\\
              $^{1}$ e.mail: \url{ramajor@prl.res.in}\\
              $^{2}$ e.mail: \url{ambastha@prl.res.in}\\
             }
%
%
%
%
\begin{abstract}
We present a technique for automatic determination of flare-ribbon separation and the energy released during the course of two-ribbon flares. We have used chromospheric H$\alpha$ filtergrams and photospheric line-of-sight magnetograms to analyse flare-ribbon separation and magnetic-field structures, respectively. Flare-ribbons were first enhanced and then extracted by the technique of ``region growing'', \ie,  a morphological operator to help resolve the flare-ribbons. Separation of flare-ribbons was then estimated from magnetic polarity reversal  line using an automatic technique implemented into Interactive Data Language (IDL\tm) platform. Finally, the rate of flare-energy release was calculated using photospheric magnetic-field data and the corresponding separation of the chromospheric H$\alpha$ flare-ribbons. This method could be applied to measure the motion of any feature of interest (\eg, intensity, magnetic, Doppler) from a given point of reference. 

\end{abstract}
\end{opening}
%
%
%

\section{Introduction}
\label{S-Introduction}
Solar flares are energetic transient events, which are produced by reconnection of magnetic-field lines at coronal heights \cite{1966Natur.211..695S,1974SoPh...34..323H,1976SoPh...50...85K}. When a coronal flux rope loses equilibrium and travels upwards, an extreme reconnection current sheet (RCS) is formed underneath. The reconnection in this RCS releases most of the magnetic energy stored in the magnetic-field configuration \cite{1984SoPh...94..315F,2000JGR...105.2375L}. Charged particles can be effectively accelerated by electric field in the RCS \cite{1990ApJS...73..333M,1995SoPh..158..317L}. Some of these energetic particles, produced during a solar flare, gyrate around the field lines and propagate toward the underlying footpoints, precipitating at different layers of the solar atmosphere to produce two bright lanes, or flare-ribbons. 
\par The \textit{Yohkoh Soft X-ray Telescope} (SXT) observed a two-ribbon structure for the first time in the hard X-ray  energy range above 30 keV, suggesting that electrons are accelerated in the whole system of a flare arcade \cite{2001SoPh..204...55M}. They analyzed the motions of two hard X-ray ribbons assuming these to be the footpoints of reconnected loops. \inlinecite{2008AdSpR..41.1195Z} studied footpoint motion of two large solar flares, including the X10 flare of 29 October 2003 using the UV/EUV observations by the \textit{Transition Region and Coronal Explorer} (TRACE) and hard X-ray (HXR) data by \textit{Reuven Ramaty High Energy Solar Spectroscopic Imager} (RHESSI). They used the 'center-of-mass' method to locate the centroids of the flare-ribbons. 
\par
\begin{figure} 
	\centering
		\includegraphics[width=1.0\textwidth]{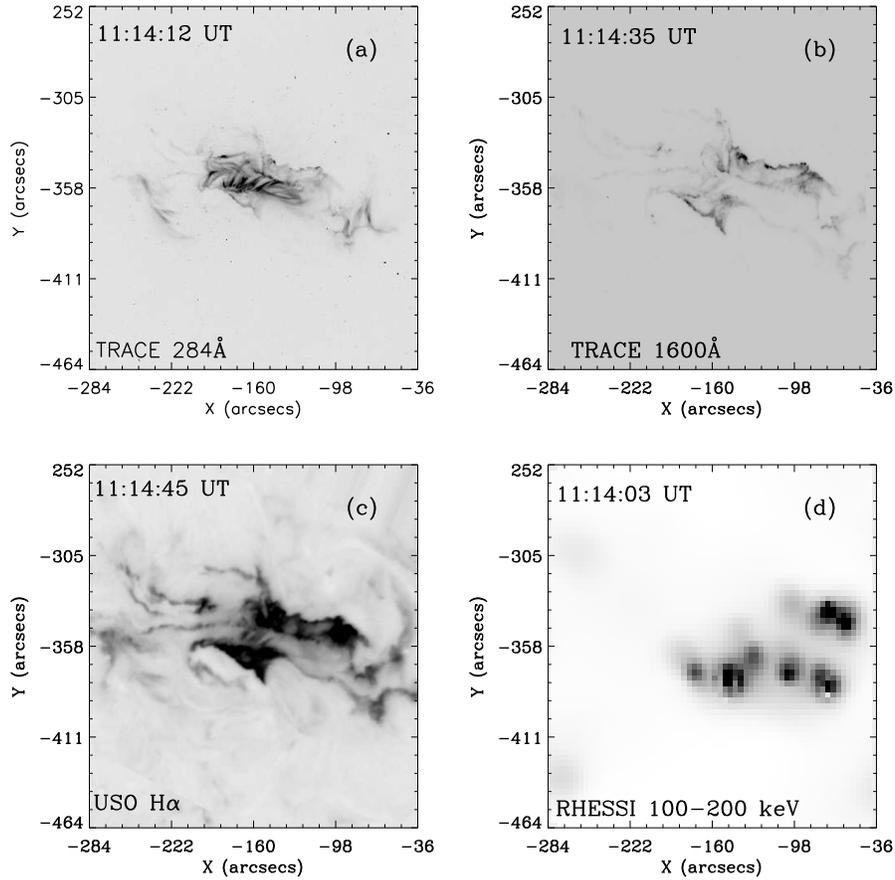}
	\caption{Images (in negative) of NOAA 10486 taken on 28 October 2003 during the X17/4B flare: (a) TRACE UV 284\,\AA~at 11:14:12 UT, (b) TRACE UV 1600\,\AA~at 11:14:35 UT, (c) USO H\,$\alpha$ at 11:14:45 UT showing post-flare loops, and (c) RHESSI HXR at 11:14:03 UT in the energy range 100\spn200 keV.}
	\label{F-tr284} 
\end{figure}

 Evolution of two\sph ribbon flares is morphologically characterized by separation of two ribbons in the chromosphere, traditionally observed in H$\alpha$. This usually occurs during solar eruptive phenomena (eruptive filaments, flares and CMEs) and is believed to be the lower atmospheric signature of magnetic reconnection progressively occurring at higher levels subsequent to energy release in the corona. As the magnetic-field lines at coronal heights progressively reconnect, their footprints move farther out from the already reconnected field lines. This successive reconnection process causes the apparent increase of the separation of flare-ribbons. As a result of magnetic reconnection, loop arcades are produced below the reconnection site. These loops are initially very hot and visible in EUV (\cf, Figure~\ref{F-tr284}a, b), and only the footpoints of the loops are observed in H$\alpha$ (\cf, Figure~\ref{F-tr284}c). The two sets of footpoints at each side of the loop arcade appear as two elongated ribbons. As the loops cool down, the whole or parts of loops may be temporarily visible in H$\alpha$. These are called H$\alpha$ (post-) flare loops. The separation velocity of flare-ribbons depends on the reconnection rate of magnetic-field lines, indicating a close relationship between flare-ribbon separation and energy release.  
\par Two well-known reconnection models are the Sweet-Parker and the Petschek models. Reconnection rate is one of the most important physical quantities in reconnection physics. \inlinecite{1958IAUS....6..123S} and \inlinecite{1957JGR....62..509P} predicted low reconnection rate due to high Reynold number ($R_m \approx10^{8}$\spn$10^{12}$) in the solar corona, and therefore the magnetic-field lines are frozen. As a result, the Sweet-Parker model is inefficient in producing fast reconnection, and conversion of magnetic energy to plasma energy takes place at a slow rate. However in a thin current sheet in the solar atmosphere, $R_m$ could be very small $(\approx 1)$ so the magnetic-field would not be frozen and could slip through plasma while the magnetic energy is converted to heat in short time scale \cite{1964NASSP..50..425P}.  This model may explain the energy-release rate and time of solar flares, and it is now widely accepted that the main driver of the solar flare is fast coronal magnetic reconnection of Petschek type.
\par The main requirement in understanding the flare reconnection rate and energy release is to detect and obtain flare characterization by image processing and pattern recognition techniques. Several methods have been proposed for automatic tracking of the apparent separation motion of two–-ribbon flares. 
\par We have developed a technique for detection of flare, its characterization and determination of various physical parameters required for calculation of the reconnection rate and energy released during the flare. Particular attention is given to the calculation of these quantities over different parts of flare-ribbons and not on the flare as a whole. We have applied our technique to the extensively observed X17/4B flare of 28 October 2003, and separation velocity of the flare-ribbons is estimated perpendicular to the magnetic neutral line. Velocities of different parts of flare-ribbons are then used to determine the energy release rate. In Section~\ref{S-formalism}, we briefly discuss the basic formalism of the problem and Section~\ref{S-Data} describes the data used in this study. Data processing steps are discussed in Section~\ref{S-proc}. The method to determine flare-ribbon separation is discussed in Section~\ref{S-ribbsep}. flare-ribbon expansion and energy release rate from different parts of the solar flare of 28 October 2003 are described in  Section~\ref{S-flare28}. Finally, the conclusions are given in Section~\ref{S-concl}.  
%
%
%
\section{The Basic Formalism}
    \label{S-formalism}
\par Properties of magnetic reconnection in the corona have been related to observed signatures of solar flares, \eg, \inlinecite{2000JASTP..62.1499F}. Their formulation requires measurement of photospheric magnetic-fields and flare-ribbon separation speeds which can be used to derive two physical terms for magnetic-reconnection rates: the rate of magnetic-flux change involved in magnetic-reconnection in the low corona and the electric field inside the RCS that is generated during magnetic-reconnection. On the basis of a reconnection model, \inlinecite{2002ApJ...566..528I} showed that the energy release rate can be written as 
\begin{equation}
  \deriv{E}{t} = S A_{\rm r} f_{\rm r}= \frac{1}{2\pi} B_{\rm c}^2 v_{\rm in} A_{\rm r} f_{\rm r}
\label{E-energy}
\end{equation}
\noindent where, $S$ is the Poynting flux into the reconnection region; $B_{c}$, $v_{\rm in}$, $A_r$ and $f_{\rm r}$ are coronal magnetic-field strength, inflow velocity, area of the reconnection region and filling factor of reconnection inflow, respectively. \inlinecite{2002ApJ...566..528I} have discussed the importance of  filling factors in the estimation of energy release rate, because not all magnetic-field lines reconnect. They suggested that it is best to use $f_r = 0.3$. However, it is difficult to measure the coronal magnetic-field and a recourse to magnetic-field extrapolation is usually made. The reconnection area can be deduced from extreme ultraviolet (EUV) images. Estimation of inflow velocity ($v_{\rm in}$) is another difficult issue, as there are very few direct observations for obtaining the inflow velocities \cite{2001ApJ...546L..69Y,2006ApJ...637.1122N}. Following an indirect method to determine the inflow velocity from magnetic-flux conservation theorem, one can write 
\begin{equation}
   B_{\rm c} v_{\rm in} = B_{\rm chro} v_{\rm ribb} =B_{\rm phot} v_{\rm ribb}
\label{E-cons}
\end{equation}
\noindent where, $B_{\rm phot}$ and $B_{\rm chro}$ are the photospheric and chromospheric magnetic-field strengths, respectively, and $v_{\rm ribb}$ is the velocity of H$\alpha$ ribbon separation. Here, we have assumed that the $B_{\rm chro} \approx B_{\rm phot}$. 
\par  The inflow velocity $v_{\rm in}$ can be thus calculated from Equation (\ref{E-cons}) using the values of $B_{\rm c}$, $B_{\rm phot}$ and $v_{\rm ribb}$. But, accurate measurement of flare-ribbon separation, $v_{\rm ribb}$, is a difficult task. Different techniques have been used to measure the separation velocity of flare-ribbons, \eg, centre-of-mass motion, label matching, \etc. \inlinecite{2004SoPh..222..137Q} developed an efficient automatic technique based on component labeling and model matching, which is useful for flare forecasting. This method, however, provides only an average separation velocity of the flare-ribbons as a whole. To overcome this limitation, we have developed a simple technique which accurately measures the direction and velocity of each component of flare-ribbons.  \inlinecite{2006JApA...27..167A} have also studied this problem in detail for the event of 10 April 2001. They identified conjugate footpoints from cross-correlation by analysing the light curve at each H$\alpha$ kernel, and tracked them to calculate  separation velocity of flare-ribbons. They estimated the energy-release rate using magnetic-reconnection model \cite{2004ApJ...611..557A} from photospheric magnetic-field and H$\alpha$ ribbon separation. They found that HXR bursts are formed around the peaks of energy-release rate.
       
%
%
%
\section{Observational Data}
    \label{S-Data}
    
We have selected a large X17/4B flare observed in NOAA 10486 on 28 October 2003 to demonstrate our technique. High spatial and temporal resolution H$\alpha$ filtergrams obtained from the Udaipur Solar Observatory (USO), Udaipur (India) have been used for obtaining the chromospheric flare-ribbon separation velocity. These filtergrams were taken by a 15-cm aperture f/15 Spar telescope at a cadence of 30 seconds during the quiet phase, and 3 seconds in the flare mode. The spatial resolution of USO filtergrams is 0.4 arcsec pixel$^{-1}$. Photospheric line-of-sight magnetograms were obtained from GONG instrument having spatial and temporal resolution of 2.5 arecsec pixel$^{-1}$ and one minute, respectively \cite{1988ESASP.286..203H}. 

\par Although we have used a rather large and complex flare in this study, the technique is useful for studying any two-ribbon flare event, in active regions or spotless regions associated with erupting filaments, which show separation with time, \eg, H$\alpha$ flare-ribbons, Doppler, and magnetic features, or for that matter, flare-ribbons observed in other wavelengths including He {\sc i} and He {\sc ii}. For example, \inlinecite{2009SoPh..258...31M} have studied the white-light flares of 28 and 29 October 2003 where flare-ribbons in multi-wavelengths, Doppler and magnetic features were analyzed. The technique can also be apply to events if we can extract the ribbons or features using mathematical morphological operations (MMO). We have listed some such events in Table~\ref{T-events}. 
\begin{table}[ht]
\caption{A list of two-ribbon flares studied using the technique}
\label{T-events}
\begin{tabular}{lcccccc}
\hline
     & Date       & Start & End   & Class   & Position & NOAA \\
     &            &       &       &         & (lat-lon)&  No. \\
\hline
1    & 28 Nov 98  & 04:54 & 06:13 & X3.3/3N &  N17E32  & 8395 \\
2    & 17 Nov 99  & 09:47 & 10:02 & M7.4    &  N18E17  & \\
3    & 17 Apr 02  & 07:46 & 09:57 & M2.6    &  S15W42  & 9906 \\
4    & 28 Oct 03  & 09:51 & 11:10 & X17.2/4B&  S16E04  & 10486 \\
5    & 29 Oct 03  & 20:37 & 21:01 & X10/2B  &  S15W02  & 10486 \\
6    & 20 Nov 03  & 07:35 & 07:53 & M9.6/2B &  N01W08  & 10501 \\
   \hline
\end{tabular}
\end{table}
  
%
%
%
%
\section{Data-Processing Technique}
    \label{S-proc}
    The raw H$\alpha$ filtergrams obtained for our study require processing and correction. There are two main image processing steps to be carried out: \textit{i}) pre-processing or data reduction and \textit{ii}) post-processing or feature development to extract the feature of interest using MMO and other techniques described in the following.
%
%
%
%
%
%
\subsection{Pre-processing or Data Reduction}
We applied the following reduction procedures: From the large set of available images for the X17/4B flare of 28 October 2003, we selected the best images taken in good ``seeing'' conditions. We applied (a) dark-subtraction to remove the signal due to dark current and (b) flat-fielding for equalization of the CCD pixels' response.

\par \textit{Image alignment (or registration)}: Effects due to random tracking errors and image motion were removed by registration of the images. This is done in two steps: First, the images were manually registered using a compact sunspot, which was assumed as a fixed reference during the period of observation. Then, a second-step registration was carried out using a Fourier technique based on a cross-correlation method on the manually registered images to get co-aligned filtergrams registration within a sub-pixel accuracy. 

\par
\begin{figure} 
	\centering
		\includegraphics[width=1.0\textwidth]{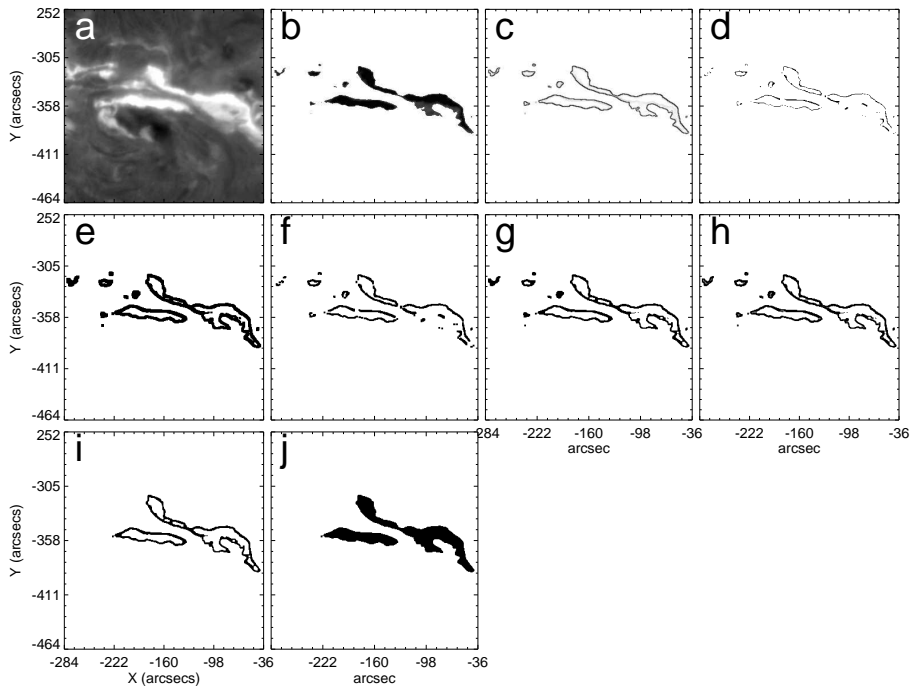}
		\caption{Results of image processing steps applied to a typical H$\alpha$ filtergram for the  X17/4B flare event of 28 October 2003 at 11:03UT: (a) image obtained after the preprocessing, (b) region-growing, (c) edge-detected using Sobel method, (d) erosion of b, (e) dilation of b, (f) opening  of b, (g) closing of b, (h) morphological closing of c, (i) small parts removed from h and (j) hole filled in i. Images b\,--\,j are obtained in negative for clarity.}
	\label{fig:im_nosaic}
\end{figure}

 \textit{Intensity level normalization (or atmospheric correction)}: The average intensity level of solar images changes with time of the day, and also due to local effects of varying observing conditions caused by dust and clouds. To remove these effects, we carried out the following correction to obtain normalized images:  
  \[
  I_{\rm out} = (I_{\rm in} -I_{\rm min}) \left (\frac{F_{\rm max}-F_{\rm min}}{I_{\rm max}-I_{\rm min}} \right ) + F_{\rm min}
  \]
  where, $I_{\rm in}$ is the corrected input image obtained from previous steps, $I_{\rm min}$ and $I_{\rm max}$ desired minimum and maximum intensity values in the output image $I_{\rm out}$, respectively. A corrected and normalized H$\alpha$ filtergram is shown in Figure~\ref{fig:im_nosaic}a. 

\par We can set value of $F_{\rm min}$ to 0 and $F_{\rm max}$ to 255 for eight-bit images. But, the problem with this is that a single outlying pixel with either a high or low value can severely affect the value of $F_{\rm min}$ or $F_{\rm max}$ and this could lead to inappropriate scaling. Therefore a better approach is to first take a histogram of the image and then select $F_{\rm min}$ and $F_{\rm max}$ at, say, the 5th and 95th percentile values in the histogram. That is, 5$\%$ of the pixel in the histogram will have values lower than $F_{\rm min}$, and 5$\%$ of the pixels will have values higher than $F_{\rm max}$. This helps in preventing the outliers pixels affecting the scaling significantly.   
%
%
%
%
%
\subsection{Post-processing or Feature Development} 
\label{SS-post} 
  
 \par For our study, we are interested in flare-ribbons, which are usually the brightest features in H$\alpha$ filtergrams. In order to select the regions of flare-ribbons in each filtergram, we set an appropriate threshold minimum intensity value. For example, a pixel value is set to zero, if it lies below 70$\%$ of the maximum value (\cf,~Figure~\ref{fig:im_nosaic}b). The threshold was decided from a plage intensity. The flare intensity is taken to be greater than this threshold. The larger threshold will lose details of flare information while smaller value will include regions lying outside the flare-ribbons. In the case of lower (upper) threshold the flare-ribbon will be thicker (thinner) but will not affect the centroid of the ribbon from where the position measurement were carried out. Therefore, the calculation of flare-ribbon separation and velocity will not be sensitive to the threshold (see, Section~\ref{S-ribbsep}). It is, however, a difficult task to decide on the boundary of flare-ribbons for which one can use boundary-based methods. We have used a Sobel edge detector adopted in IDL\tm. Edges of flare-ribbons in Figure~\ref{fig:im_nosaic}b, detected using this method, are shown in Figure~\ref{fig:im_nosaic}c. 

\par After obtaining the pre-processed images, we proceed to enhance and extract features of interest, \eg, flare-ribbons, using MMO, which is a powerful tool to extract the main feature of a digitized image \cite{2008dip..book.....G}. In image processing, this tool was used for the first time by  \inlinecite{1983dip..book.....S}  to find the geometrical structures in images. More recently, it has been used in solar physics for detection of flares (\citeauthor{2004SoPh..222..137Q} \citeyear{2004AAS...204.5416Q,2004SoPh..222..137Q}), filaments \cite{2003SoPh..218...99S,2005SoPh..228..119Q,2008Angeo...26..243A}, sunspots (\citeauthor{2005Eu..2005..2573Z} \citeyear{2005Eu..2005..2573Z,2005Eu..23..209Z})
 and prominences \cite{2007IEEE...17....1F}.  Another technique based on neural networks has also been used for the detection of solar flares by Fernandez Borda \etal~\shortcite{2002SoPh..206..347F}. 

\par The language of MMO is based on set theory. Sets in mathematical morphology represent objects in an image. A binary image is a complete morphological description of the image in 2D integer space ($\rm{Z^2}$), where elements are either ``0'' or ``1''. The fundamental MMO are ``erosion'' and ``dilation''. If $A$ and $B$ are sets in $\rm{Z^2}$ space then, ``erosion'' of $A$ by $B$ is defined as 
\begin{equation}
	A\ominus B = \{z: (B)_z \subseteq A \}
	\label{E-eros}
\end{equation}
In this study, A is the image (e.g., Figure 2b) to be eroded and B is the 3$\times$3 structuring element.  In erosion, smaller size of structuring element will eliminate smaller components while larger size will eliminate larger components. Larger size will deform the shape of flare-ribbons which will change their centroid. This may affect the measurements of flare-ribbon separation and velocity. A more description about the structuring element with different MMO can be found in \inlinecite{2008dip..book.....G}.

\par The ``erosion'' shrinks the components of an image. It consists of replacing each pixel of an image by the minimum of its neighbours (Figure~\ref{fig:im_nosaic}d). The ``dilation'' of $A$ by $B$ in $\rm{Z^2}$ space is defined as 
\begin{equation}
  	A\oplus B = \{z: (\hat{B})_z \cap A \ne \phi \}
	\label{E-dila}
\end{equation}
The ``dilation'' expands the components of an image. It consists of replacing each pixel of an image by the maximum of its neighbours (Figure~\ref{fig:im_nosaic}e).  
There are two other morphological operators: ``opening'' and ``closing''. The ``opening'' generally smooths the contour of an object, breaks narrow isthmuses, and eliminates thin protrusions. The ``closing'' also tends to smooth sections of contours but, as opposed to ``opening'', it generally fuses narrow breaks and long thin gulfs, eliminates small holes, and fills gaps in the contour. The ``opening'' of a set A by structuring element $B$ is defined as
\begin{equation}
  	A \circ B = (A\ominus B)\oplus B
	\label{E-open}
\end{equation}
\noindent \ie, opening of $A$ by $B$ is the erosion of $A$ by $B$, followed by a dilation of the result by $B$, Figure~\ref{fig:im_nosaic}f. Similarly, ``closing'' of set $A$ by structuring element $B$ is defined as
\begin{equation}
  	A \bullet B = (A\oplus B)\ominus B
	\label{E-close}
\end{equation}
\noindent \ie, closing of $A$ by $B$ is the dilation of $A$ by $B$, followed by an erosion of the result by $B$ (Figure~\ref{fig:im_nosaic}g).
We applied morphological ``closing'' to the image (\ref{fig:im_nosaic}c) to erase the gaps and smooth the contours. A binary image after this operation is shown in Figure~\ref{fig:im_nosaic}h. The images after applying MMO consist of many small features. These are removed using ``region labeling'' methods.  The region labeling method gives labels to each components of an image. We count the number of pixels in each label and remove those component by ``0'', which have values less than a set threshold (Figure~\ref{fig:im_nosaic}i).  

\par After removing small components from images obtained using the above method, we find small holes, or background regions surrounded by a connected border of foreground pixels within selected components, \ie, large area components as seen in Figure~\ref{fig:im_nosaic}i. In fact, the holes in between the flare-ribbons are part of the flare-ribbons, created due to edge detection by the Sobel operator. Since we are interested only in the edge, area, etc. of flare-ribbons, these holes should be removed by ``1'' from binary images. For this purpose, we have used an automatic procedure based on morphological reconstruction \cite{2008dip..book.....G}. This procedure uses ``geodesic dilation''. This is described as follows. 

\par Let $I(x,y)$ be a binary image. We form a marker image F, which is ``0'' everywhere except at the image border, where it is set to $1-I$; \ie,
\begin{equation}
F(x,y) = \left\{ \begin{array}{l}
 1 - I(x,y)\hspace{0.2cm} \textrm{if}~(x,y)~\textrm{is on the border of I} \\ 
 0 \hspace{1.7cm}{\rm otherwise} \\ 
 \end{array} \right.
\end{equation}
Then, the binary image with holes filled will be,
\begin{equation}
H = \left[ {R_{I^c }^n (F)} \right]^c.
\end{equation}
Superscript $c$ represents the complement and $\left[ {R_{I^c }^N (F)} \right]^c $ is ``geodesic dilation'' of size $n$ of the marker image $F$ with respect to the mask image $I^c$. The ``geodesic dilation'' is defined by,
\begin{equation}
\left[ {R_{I^c }^n (F)} \right]  = R_{I^c}^{(1)}\left[{R_{I^c}^{(n-1)} (F)} \right] 
\end{equation}
where,
$R_{I^c}^{(1)}(F)=(F\oplus B)\cap I^c$.
$\cap$ represents the ``intersection'' operator and B is the structuring element as described above. Image after removing the holes is shown in Figure~\ref{fig:im_nosaic}j.

%
%
\section{Determination of the Flare-ribbon Separation}
 \label{S-ribbsep}
 Once the flare-ribbons are extracted, we can estimate their separation from a refrence, such as the magnetic neutral line. Some questions that arise in the determination of flare-ribbon separation are: Can we use a straight line representing the neutral line passing between the flare-ribbons? Do the flare-ribbons move perpendicularly to the neutral line? Do the flare-ribbons on either side move with the same velocity? From the visual inspection of the H$\alpha$ flare-ribbons (\cf,~Figure~\ref{fig:im_nosaic}a), we find that the ribbons are nearly parallel to each other, but they are curved in shape. Therefore, we can not draw a straight line representing the neutral line. Also, various components of the flare-ribbons are curved over different directions. Therefore, the direction of neutral line is ``important'' to get the direction of motion of a given part of the flare-ribbon. 
\begin{figure}[ht] 
	\centering
		\includegraphics[width=0.6\textwidth]{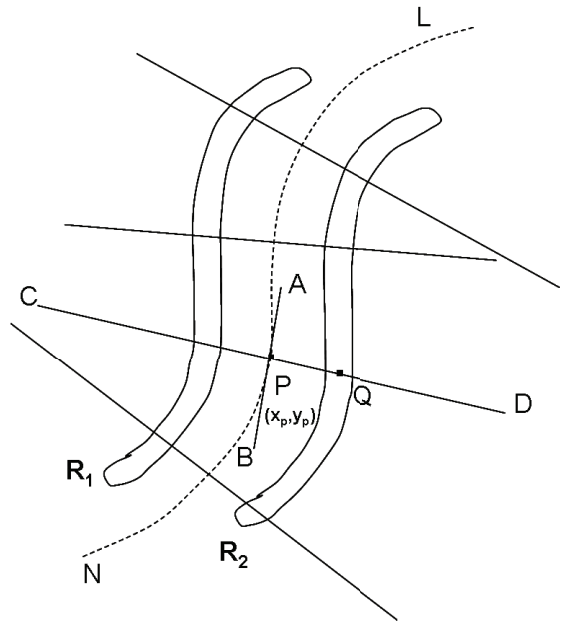}
		\caption{Cartoon of a two-ribbon flare. The dotted line represents the magnetic polarity  reversal line. The two contours around the dotted line represent the two ribbons $R_1$ and $R_2$ of the flare. The solid line AB is tangent to the neutral line NL at point P. Other solid lines are drawn perpendicular to the neutral line at different points.}
	\label{fig:ribcort}
\end{figure}
 \par Figure \ref{fig:ribcort} shows a cartoon of two ribbon flare to illustrate the method used here for distance measurement between flare-ribbons and the neutral line (NL). $R_1$ and $R_2$ represent two ribbons of the flare, and the dotted line NL represents the magnetic neutral line. Solid lines, drawn perpendicular to the neutral line, mark the directions of motion of different portions of ribbons, assuming that the apparent motion of flare-ribbons is perpendicular to the neutral line. Therefore, the velocity derived for a given point on the flare-ribbon may be the same in magnitude, but it may not necessarily be so in direction. It is important to consider this issue appropriately in the analysis. Further, it is observationally known that the velocity of flare-ribbons is affected by the strength of the magnetic-field in the region of flare-ribbon motion. However, it should be noted that this is only an apparent motion which may not represent the real motion of the flare arcade.

\par It is well-known that the overlying arcade has shear, which decreases with increasing distance from the magnetic-polarity reversal boundary and with increasing coronal height. Thus it follows that footpoints of the reconnected flare loops will move not only from the polarity reversal boundary, but also along it. To address this issue, we need to obtain 3D topology of the magnetic-reconnection region. We should also know the conjugate points joining the flare-ribbons, which is not easy. Some information about this can be obtained from magnetic-field extrapolation and post-flare loops seen in EUV images. For example, to find the conjugate points in H$\alpha$ flare-ribbons, \inlinecite{2006JApA...27..167A} divided the regions of flare-ribbons lying over magnetic polarities opposite to each other into fine meshes. Then, they drew light curves of total intensity for each box in both meshes, and identified the highly correlated pairs (conjugated pairs) by calculating the cross-correlation functions. This is a good approach to find the conjugate points, but for a complex region, field lines may be highly sheared in small scales which poses difficulty in identifying these points. The magnetic-field extrapolation may be a more accurate approach to get much closer to the conjugate pairs. But, again, for complex regions errors in magnetic-field extrapolation pose difficulty to identify the exact conjugate points. Therefore, we have taken a simple assumption of flare-ribbon motion perpendicularly to the neutral line.  

\par In view of the above-mentioned issues, we have implemented an algorithm in IDL\tm~for detection of flare-ribbon separation based on an automatic technique. It calculates the ribbon separation measured from the neutral line at specified points of the neutral line. The algorithm assumes that the ribbons move perpendicularly to the neutral line. The main steps of the technique are:
 
\begin{itemize}
	\item \textit{Neutral line passing between the flare-ribbons}: We overlaid contours of co-spatial photospheric line-of-sight (LOS) magnetograms over the corresponding H$\alpha$ filtergrams observed around the time of the flare. The neutral line (NL) drawn in green colour is obtained by fitting an appropriate polynomial over several points within the region of magnetic-field lines drawn at $\pm$5 Gauss levels (Figure~\ref{fig:neut_line}, left panel).  
\begin{figure}[ht] 
	\centering
	  \includegraphics[width=0.49\textwidth,clip=,bb=53 53 306 306]{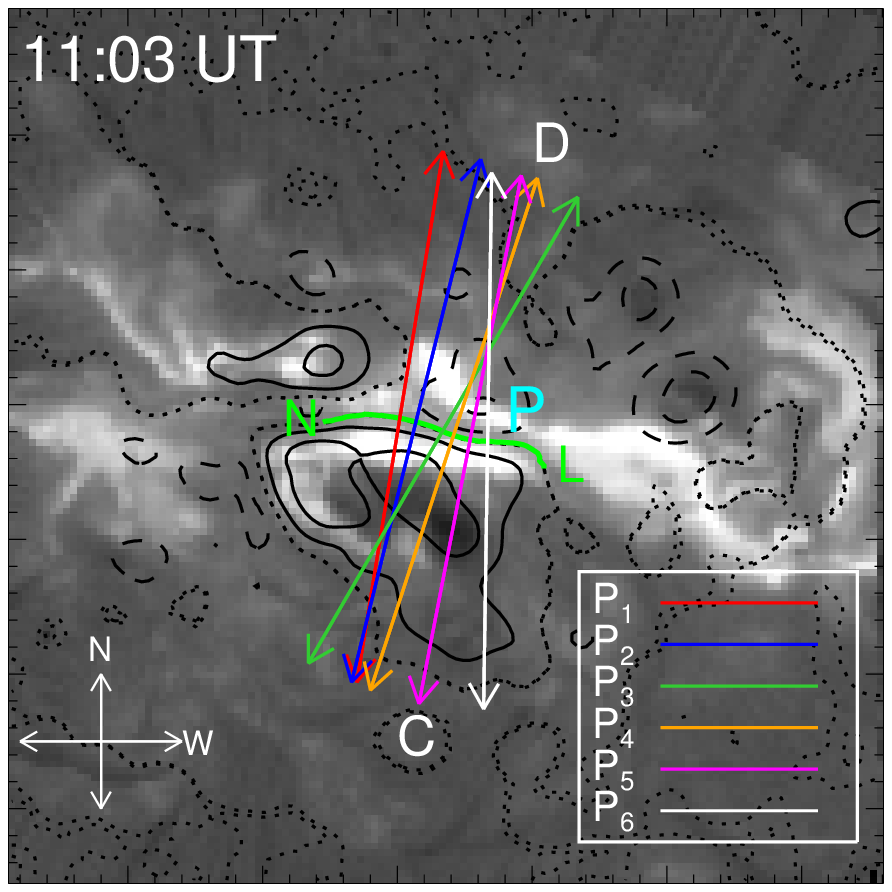} 
	  \includegraphics[width=0.49\textwidth,clip=,bb=36 36 323 323]{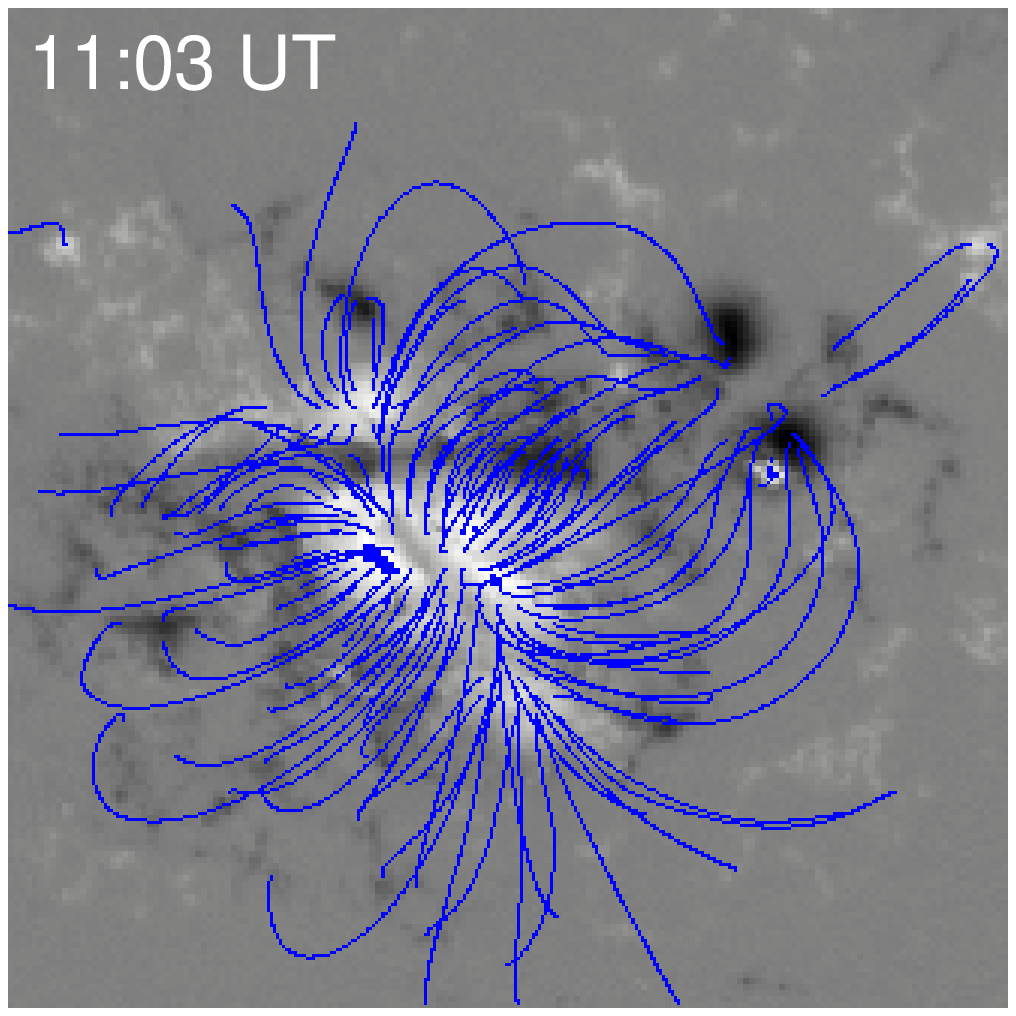}
	  \caption{Left panel: H$\alpha$ filtergram of NOAA 10486 overlaid with the contours of longitudinal component of magnetic-field showing the  X17/4B flare of 28 October 2003 at 11:03 UT. Solid and dashed contours  represent positive and negative polarities, respectively. Dotted curves represent the polarity reversal lines. A part of neutral line NL passing through the flare-ribbons is highlighted in green colour. Right panel: Extrapolated magnetic-field lines plotted over the GONG LOS magnetogram using the IDL package MAGPACK2.}
	\label{fig:neut_line}
\end{figure}

 \item \textit{Direction of motion of the flare-ribbons}: For simplicity, we assume that flare-ribbons move perpendicularly to the neutral line. Let NL be the neutral line (\cf,~Figure~\ref{fig:ribcort}) passing through the flare-ribbons $R_1$ and $R_2$, and P be a point on it. AB is tangent at P on NL, and CD is perpendicular to AB at P. Let Q be the centroid of the part of flare-ribbon $R_2$ which lie along PD and follows path PD. It is required to find the direction of PD with respect to the horizontal or x-axis. For this, we find the gradient (say, $m_p$) at point P derived by Lagrange interpolation method. Therefore, the gradient of line PD will be $m'_p=-1/m_p$. A similar procedure can be carried out to find the direction of motion of a flare-ribbon at any desired point along NL. To obtain the coordinates of points on a line, say PD, we have used a simple algorithm as described in the Appendix. 

	 \item \textit{Distance $d_t$ of a point Q($x_k,y_k$) on the flare-ribbon} at time $t$ from point $P(x_p,y_p$) on the neutral line is given by  
	        \[
	        d_t=\sqrt{(x_k-x_p)^2 + (y_k-y_p)^2}
	        \]

and the direction of motion is
	    \[
	    \theta_p=\mathrm tan^{-1} (m^\prime_p)
	    \]      
\end{itemize}
 
Using the calculated flare-ribbon distances from the neutral line,  one can find the velocity simply by taking their time derivatives. But this would be noisy due to the errors in the measured distances using above procedures. Therefore, an appropriate function is used to fit the measured distances so that the errors are reduced.

\begin{figure} 
\centering
\begin{minipage}[t]{0.5\textwidth}
\centering
\includegraphics[width=1.0\textwidth]{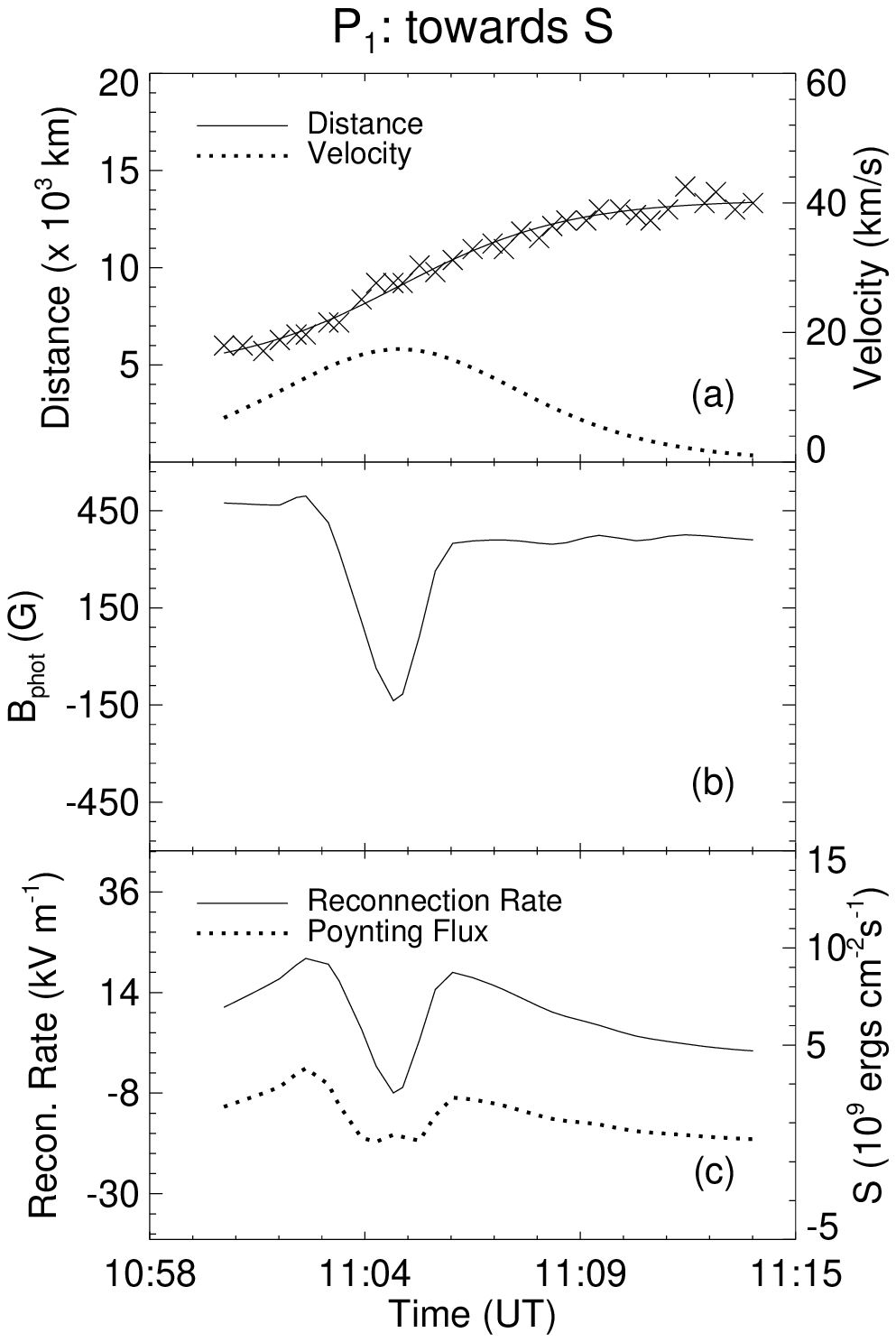}
\end{minipage}%
\begin{minipage}[t]{0.5\textwidth}
\centering
\includegraphics[width=1.0\textwidth]{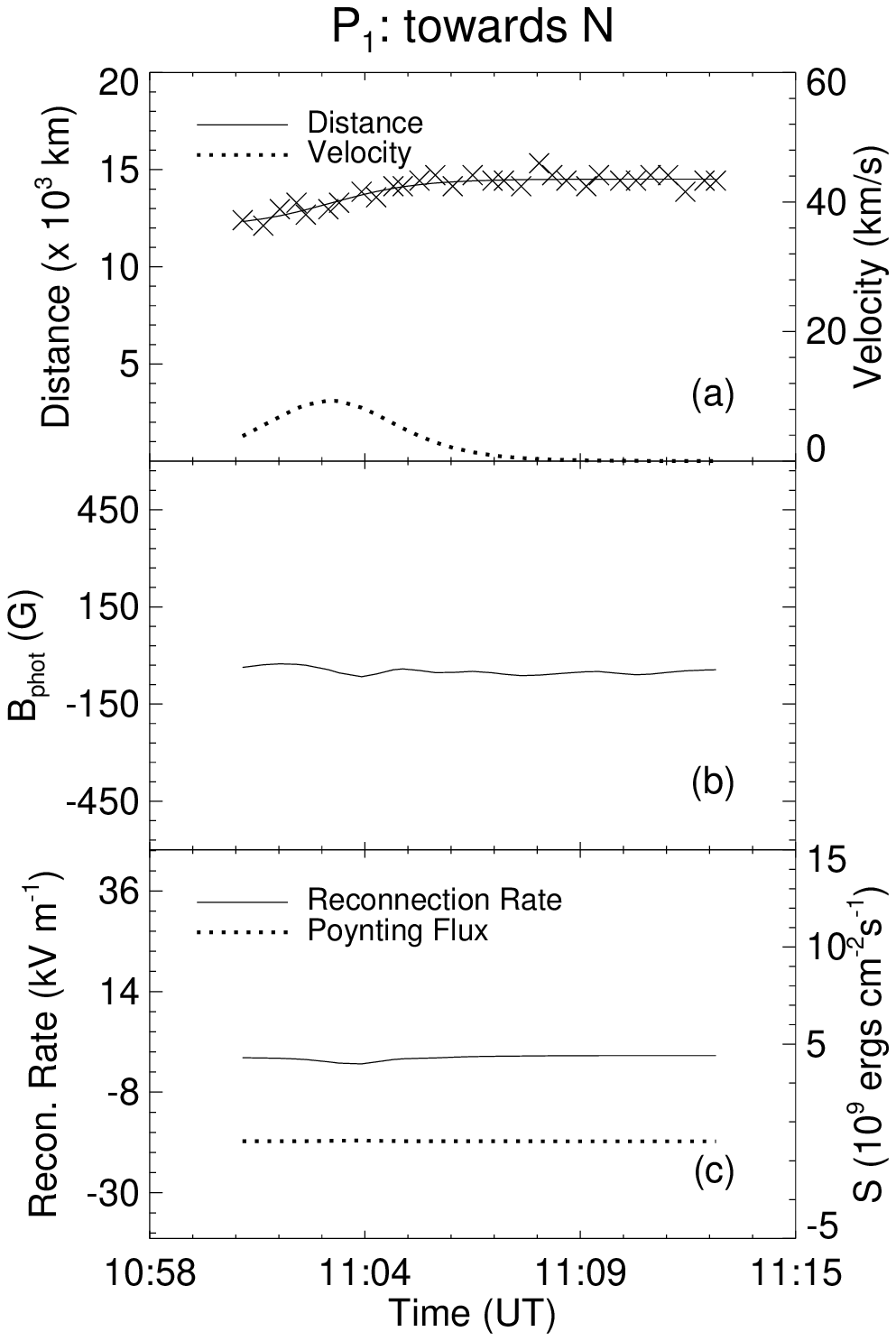}
\end{minipage}
\centering
\begin{minipage}[t]{0.5\textwidth}
\centering
\includegraphics[width=1.0\textwidth]{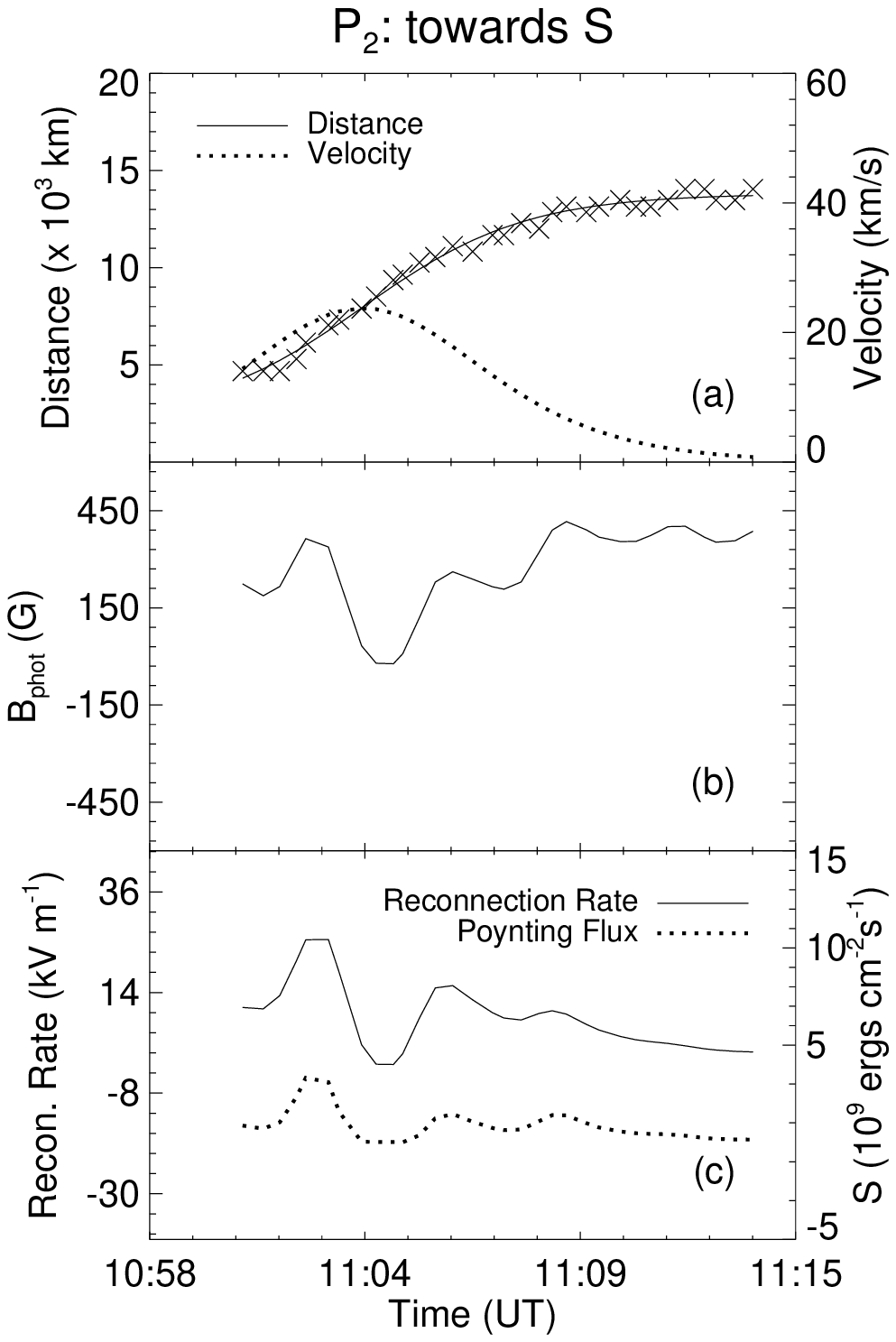}
\end{minipage}%
\begin{minipage}[t]{0.5\textwidth}
\centering
\includegraphics[width=1.0\textwidth]{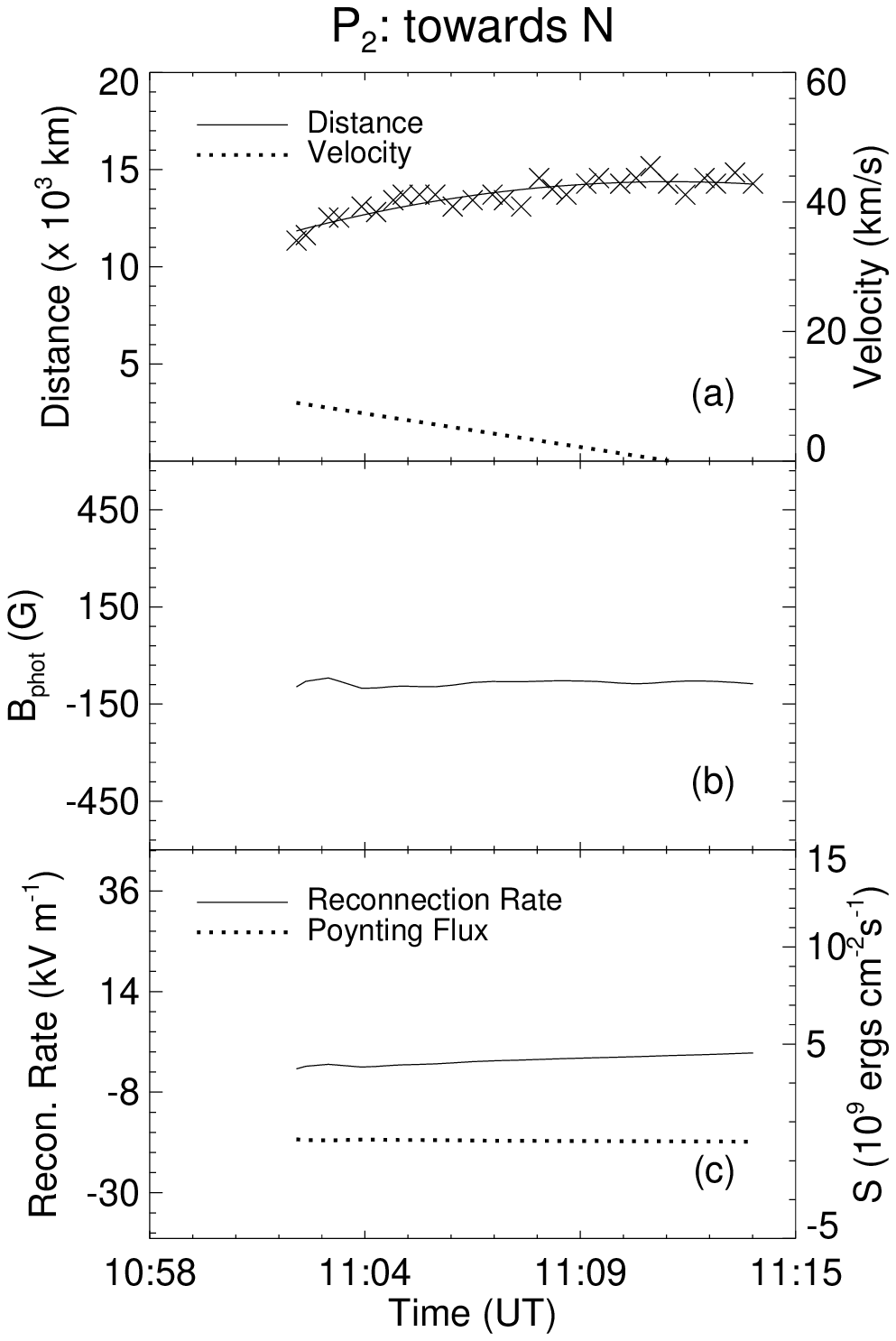}
\end{minipage}
	\caption{Top and bottom panels correspond to the measurement of different parameters along the lines $P_1$ and $P_2$,  respectively (\cf, Figure~\ref{fig:neut_line}). (a) `'$\times$'' represents the measured distances, and the solid curves passing through these are the Boltzmann--Sigmoid fitted distance profiles with time. Dotted lines represent the velocity $v(t)$ profiles derived by taking time derivative of the fitted distances. (b) Magnetic-flux $B_{\rm phot}$ at the points used for distance measurement. (c) Reconnection rate ($\dot{\Phi}$) (solid line) and Poynting flux ($S$) (dotted lines).}
		\label{fig:dvmrp1}
\end{figure}
\begin{figure} 
\centering
\begin{minipage}[t]{0.5\textwidth}
\centering
\includegraphics[width=1.0\textwidth]{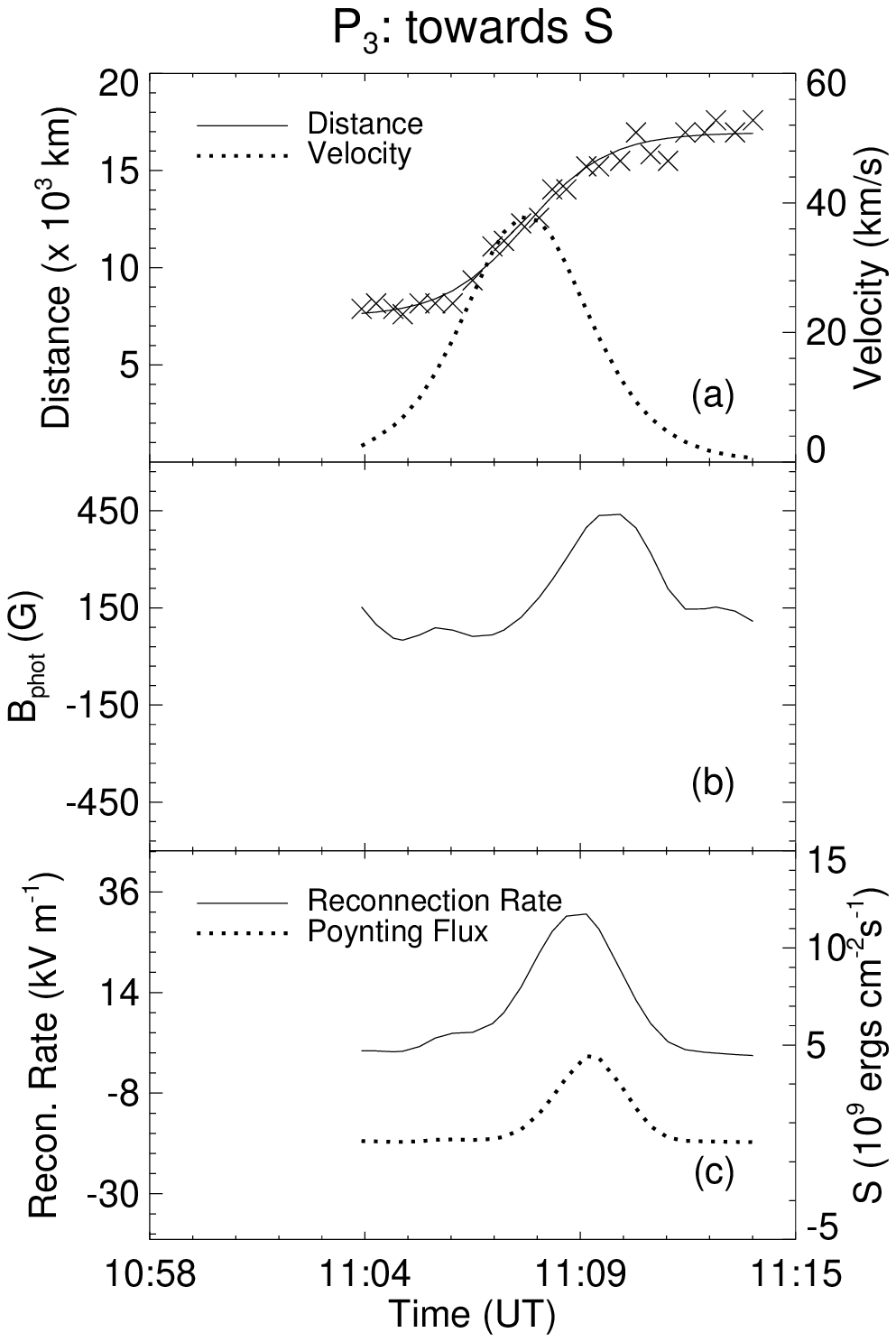}
\end{minipage}%
\begin{minipage}[t]{0.5\textwidth}
\centering
\includegraphics[width=1.0\textwidth]{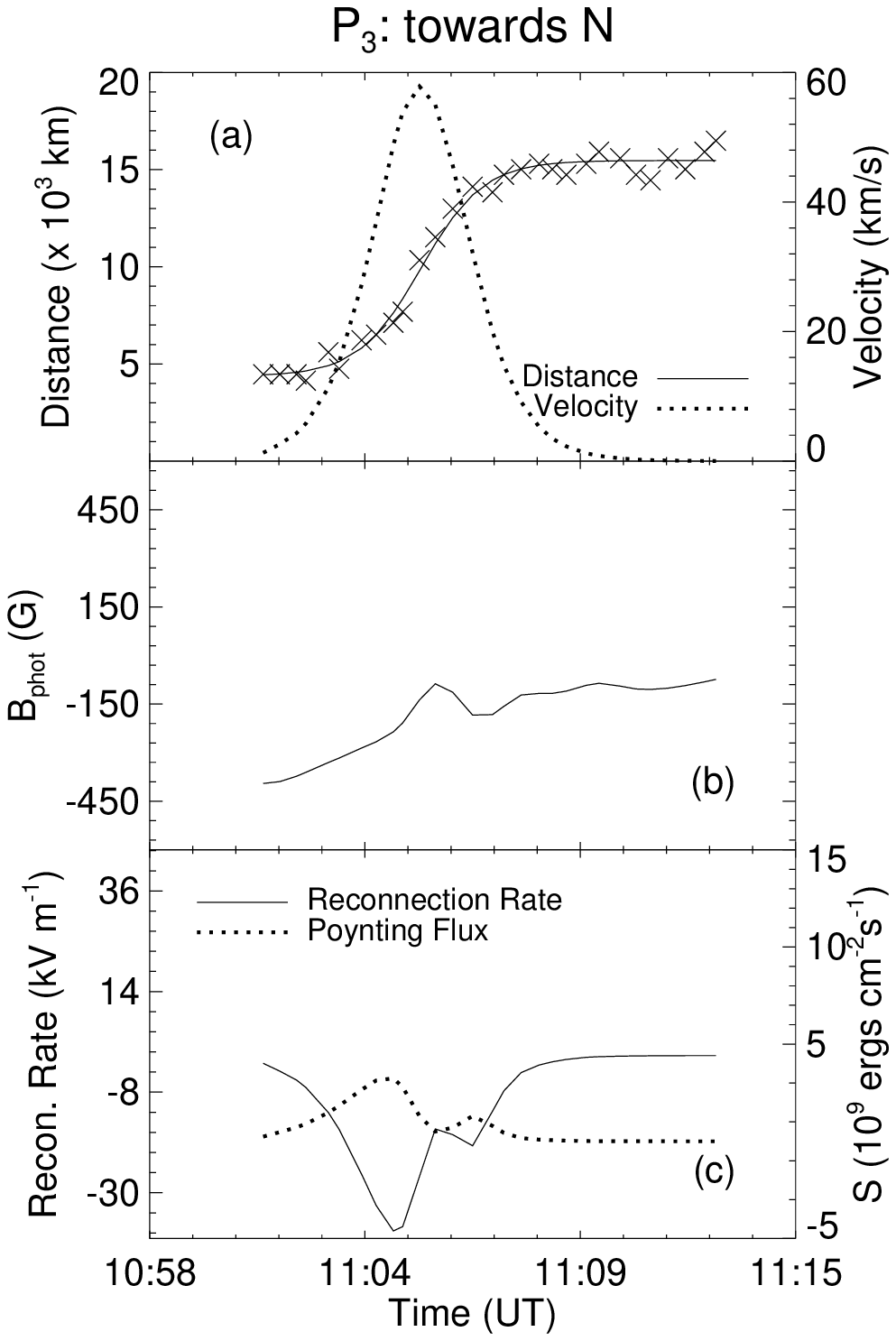}
\end{minipage}
\centering
\begin{minipage}[t]{0.5\textwidth}
\centering
\includegraphics[width=1.0\textwidth]{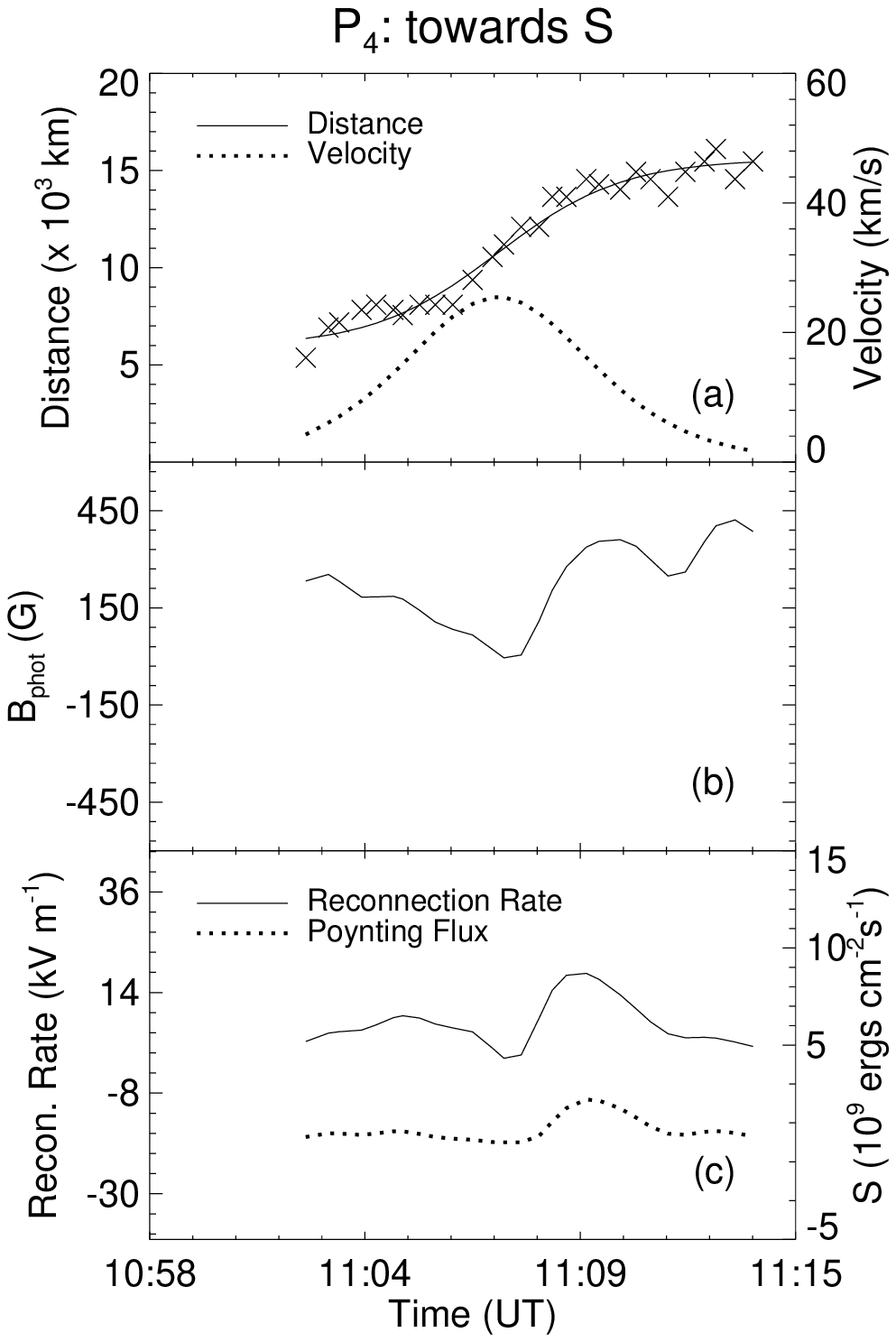}
\end{minipage}%
\begin{minipage}[t]{0.5\textwidth}
\centering
\includegraphics[width=1.0\textwidth]{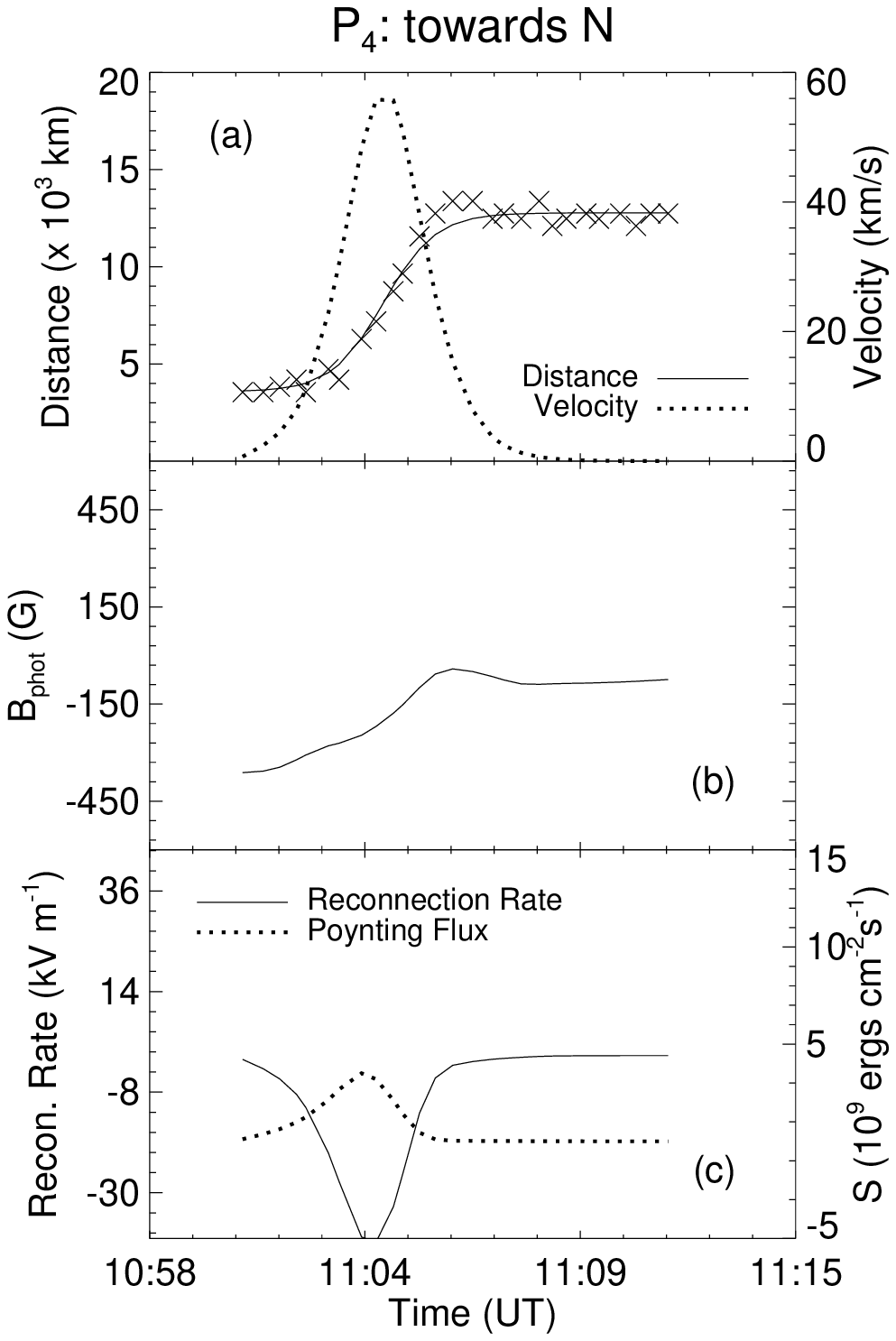}
\end{minipage}
\caption{Top and bottom panels correspond to the measurement of different parameters along the lines $P_3$ and $P_4$,  respectively (\cf, Figure~\ref{fig:neut_line}). (a) ``$\times$'' represents the measured distances, and the solid curves passing through these are the Boltzmann--Sigmoid fitted distance profiles with time. Dotted lines represent the velocity $v(t)$ profiles derived by taking time derivative of the fitted distances. (b) Magnetic-flux $B_{\rm phot}$ at the points used for distance measurement. (c) Reconnection rate ($\dot{\Phi}$) (solid line) and Poynting flux ($S$) (dotted lines).}
		\label{fig:dvmrp3}
\end{figure}
\begin{figure} 
\centering
\begin{minipage}[t]{0.5\textwidth}
\centering
\includegraphics[width=1.0\textwidth]{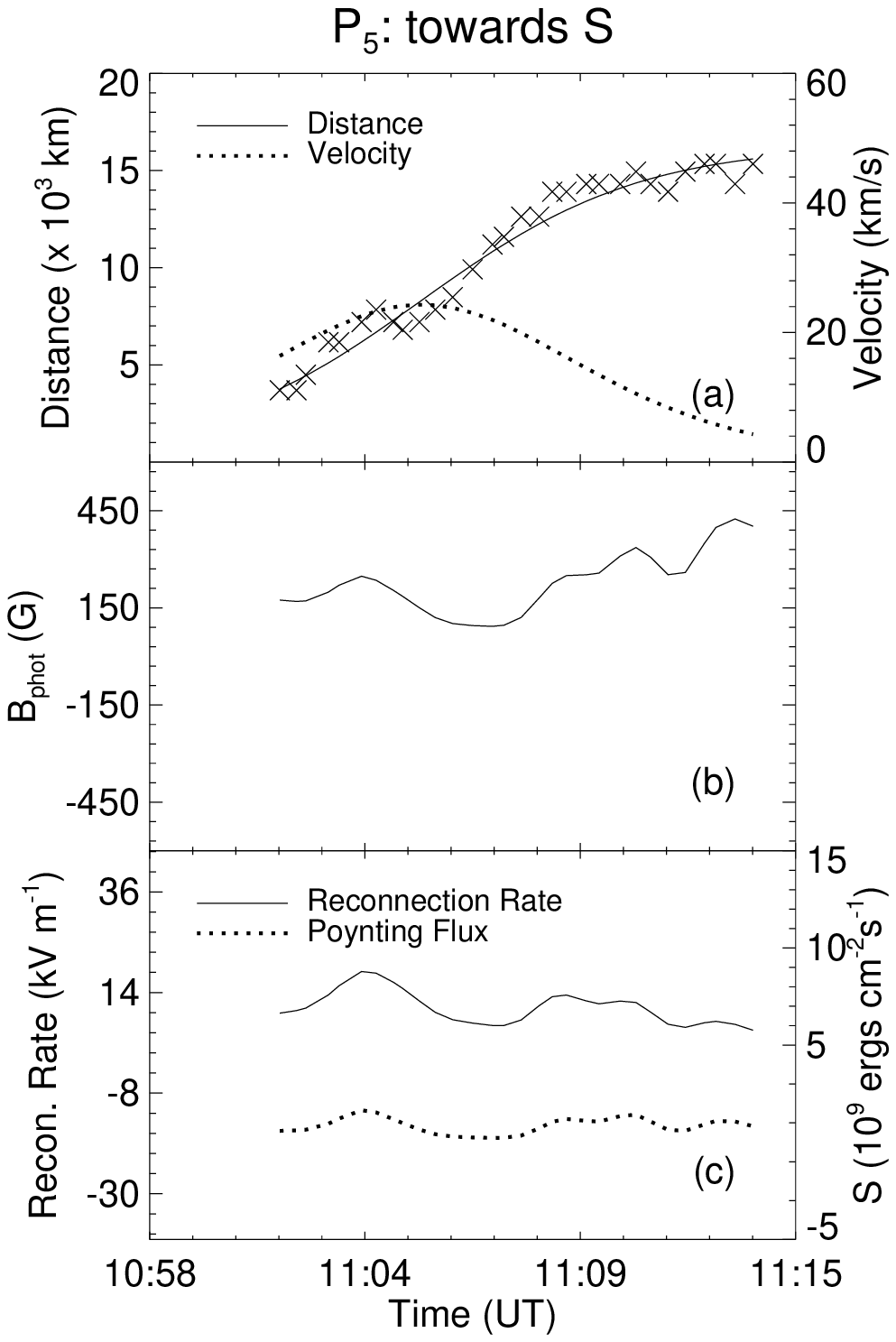}
\end{minipage}%
\begin{minipage}[t]{0.5\textwidth}
\centering
\includegraphics[width=1.0\textwidth]{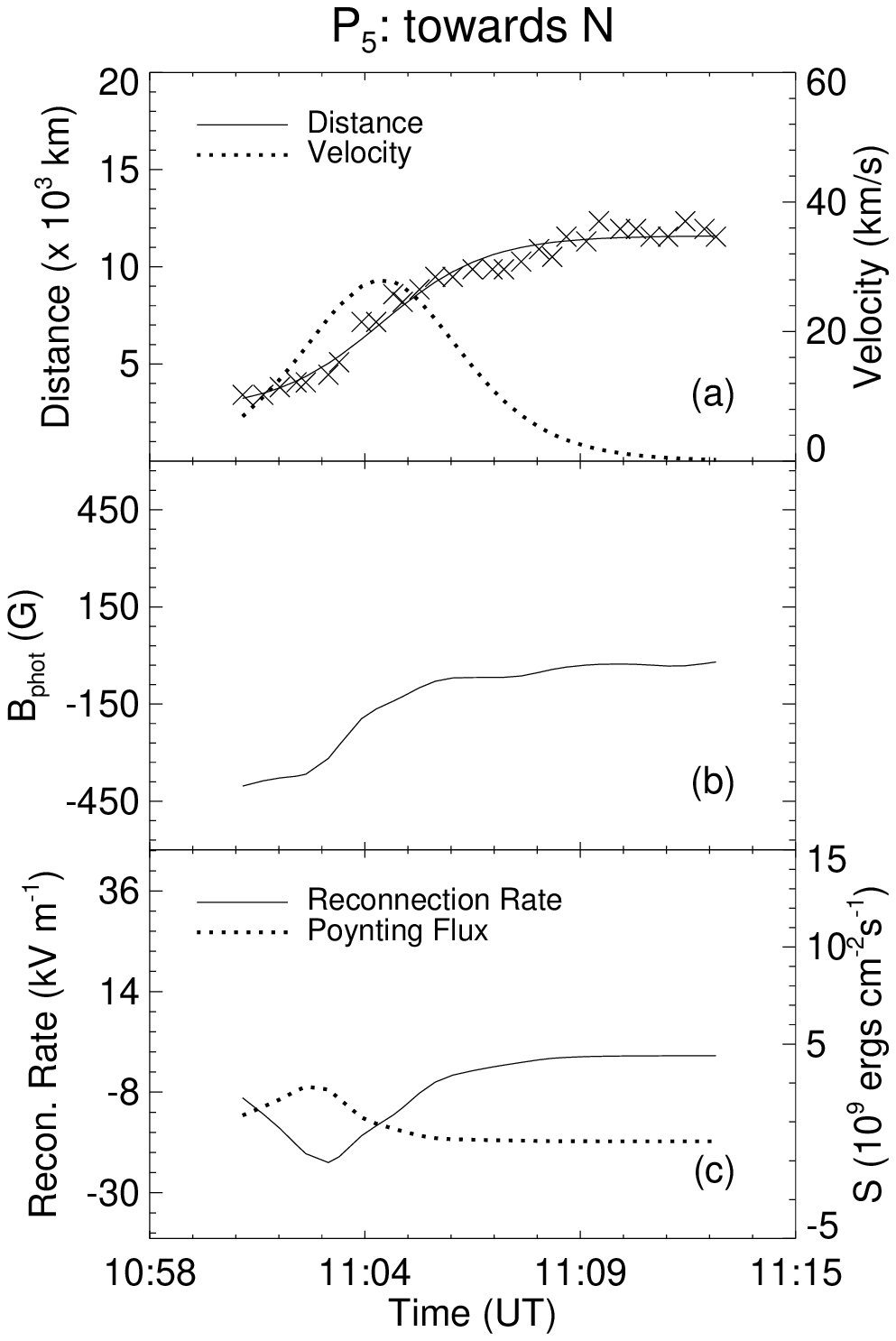}
\end{minipage}
\centering
\begin{minipage}[t]{0.5\textwidth}
\centering
\includegraphics[width=1.0\textwidth]{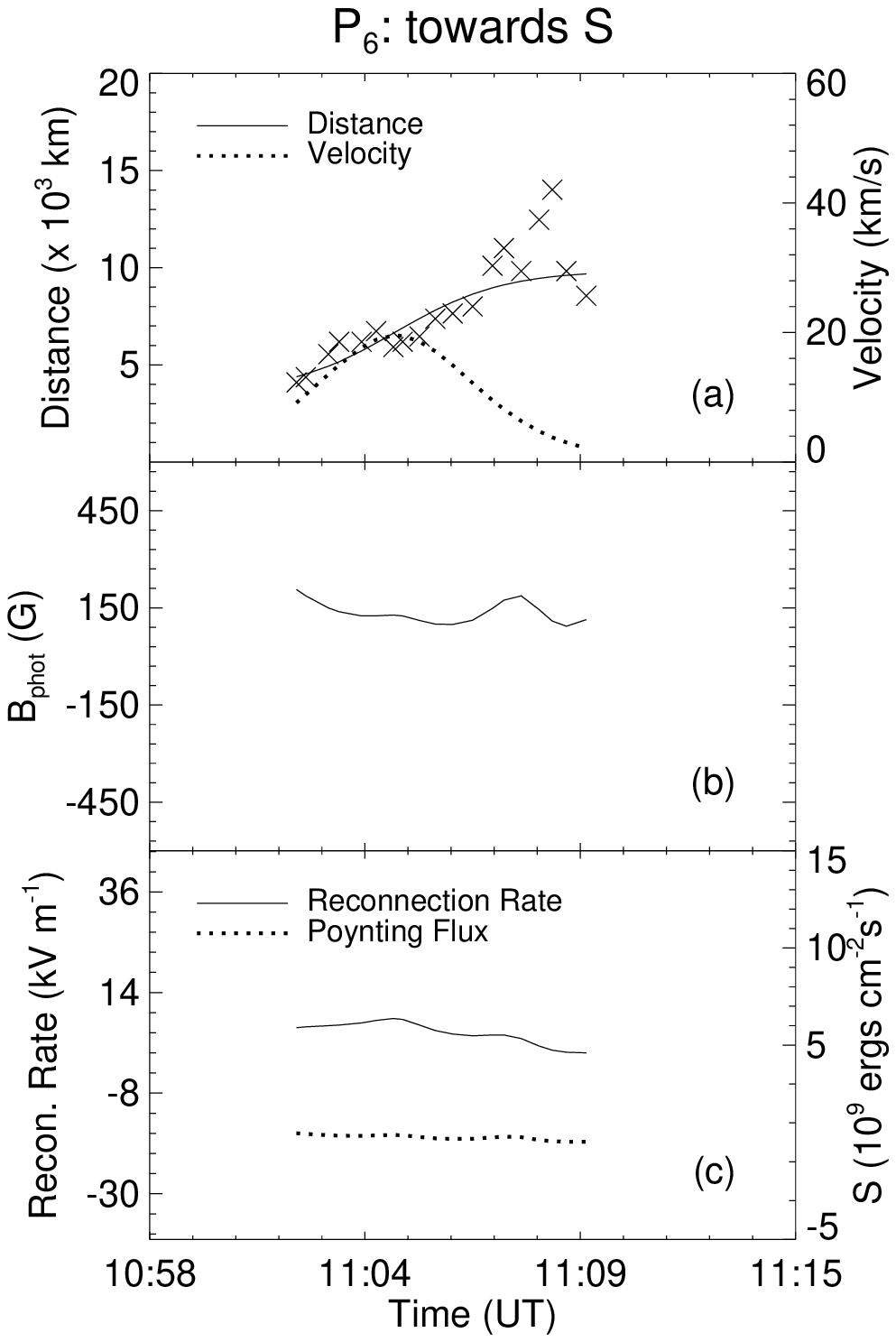}
\end{minipage}%
\begin{minipage}[t]{0.5\textwidth}
\centering
\includegraphics[width=1.0\textwidth]{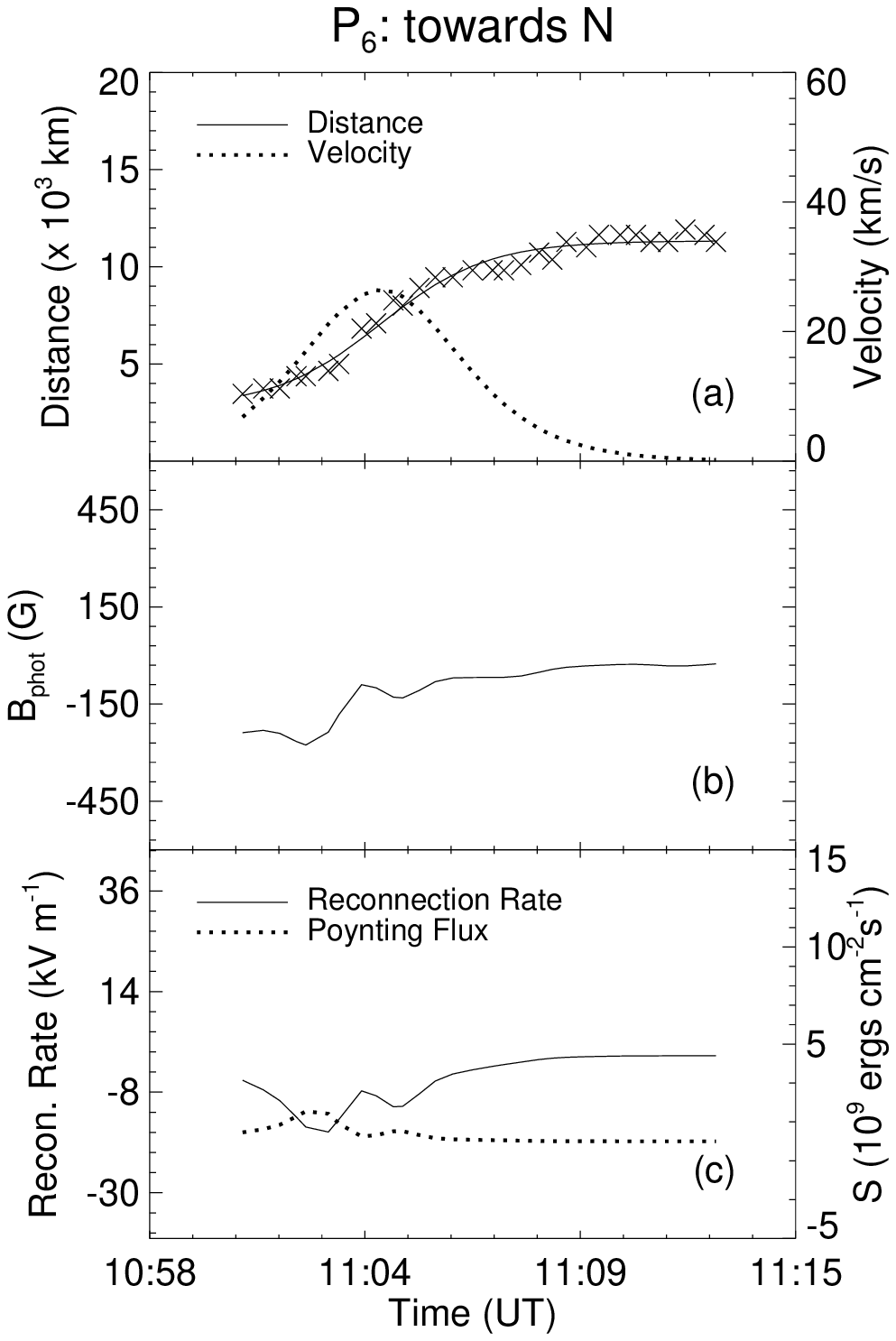}
\end{minipage}
\caption{Top and bottom panels correspond to the measurement of different parameters along the lines $P_5$ and $P_6$,  respectively (\cf, Figure~\ref{fig:neut_line}). (a) ``$\times$'' represents the measured distances, and the solid curves passing through these are the Boltzmann--Sigmoid fitted distance profiles with time. Dotted lines represent the velocity $v(t)$ profiles derived by taking time derivative of the fitted distances. (b) Magnetic-flux $B_{\rm phot}$ at the points used for distance measurement. (c) Reconnection rate ($\dot{\Phi}$) (solid line) and Poynting flux ($S$) (dotted lines).}
		\label{fig:dvmrp6}
\end{figure}
%
%
%
%
%
\section{H$\alpha$ Flare-ribbon Separation and Energy-release}
\label{S-flare28}

The X17/4B event of 28 October 2003 was a two-ribbon white light flare that occurred in the super active region NOAA 10486. An H$\alpha$ filtergram for this event at 11:03 UT overlaid with contours of longitudinal magnetic-field shown in Figure~\ref{fig:neut_line}a. Solid and dashed lines  represent positive and negative magnetic polarities, respectively, and the dotted lines represent magnetic neutral lines. The neutral line (NL) passing through the flare-ribbons is highlighted in green. Straight lines drawn perpendicular to the neutral line mark the directions of motion of different parts of flare-ribbons from the corresponding points $P_{\rm i}\,(\rm i = 1,\,\ldots,6)$ of the neutral line. Here, we have selected six points $P_i$ over the neutral line and corresponding directions of flare-ribbon motion are shown. The center of the two edges of a flare-ribbon is followed along the line passing through $P_{\rm i}$ to measure the flare-ribbon separation from corresponding points (\ie, $P_{\rm i}$) on NL. The measured distances corresponding to different components are shown by ``$\times$'' symbol in the panel (a) of Figures (\ref{fig:dvmrp1})\spn(\ref{fig:dvmrp6}).
 
 \par Velocities corresponding to six selected parts of the flare-ribbon were obtained by taking time derivatives of the fitted distances. We found Boltzmann--Sigmoid to be the best fit for this case \cite{2009SoPh..258...31M}. The fitted distances are shown by solid lines passing through the measured points in the panel (a) of Figures (\ref{fig:dvmrp1})\spn(\ref{fig:dvmrp6}). Corresponding velocities are shown by dotted lines in the same panels. 
%
%
%
%
%
\subsection{Magnetic-reconnection and Energy-release}
 \label{Sb-energy}

Using the derived separation velocities for the flare-ribbons, we can estimate magnetic-reconnection rate and energy released during the solar flare. Magnetic- reconnection theories predict how fast reconnection can occur. The speed of magnetic-reconnection can be expressed in terms of the inflow velocity $v_{\rm in}$ or the dimensionless ratio of $v_{\rm in}$ to the Alfv$\rm{\acute{e}}$n velocity ($v_{\rm A}$) in the inflow region.

\[
M_i = \frac{v_{\rm in}}{v_A}
\]

\noindent where, the Alfv$\rm{\acute{e}}$n velocity is given by $v_A = B_{\rm c}/ \sqrt{4 \pi \rho}$ and $\rho$ is density near the reconnection region, which can be taken as a free parameter from an atmospheric model. Therefore, the reconnection rate using Equation (\ref{E-cons}) can be written as,

\begin{equation}
  M_i = \frac{v_{\rm ribb} B_{\rm phot} \sqrt{4 \pi \rho}}{B_{\rm c}^2 }
	\label{eq:recr}
\end{equation}
\noindent An alternative measure of the reconnection rate is the electric-field strength in the RCS. It shows how violently the magnetic-reconnection progresses. It is defined as the reconnected magnetic-flux per unit time and is expressed as, 
\[
 \dot{\Phi} = B_{\rm c} v_{\rm in}
\]
\noindent From Equation (\ref{E-cons}), it can be written as
\begin{equation}
   \dot{\Phi} = B_{\rm phot} v_{\rm ribb}
  \label{E-rec_rate}
\end{equation}  
\noindent The rate of magnetic energy-release, \ie, energy released per unit time during a solar flare, is the product of Poynting flux and the area of RCS, given by Equation (\ref{E-energy}). If we assume that the area of magnetic-reconnection is fixed during the flare (as assumed by \opencite{2004ApJ...611..557A}), the energy-release rate will be proportional to the Poynting flux only and hence independent of the filling factor $f_{\rm r}$. Using Equations (\ref{E-energy}) and (\ref{E-cons}), Poynting flux can be written as,

\begin{equation}
   S=\frac{1}{2\pi}B_{\rm c} B_{\rm phot} v_{\rm ribb}
\label{E-poy1}
\end{equation}
     
\noindent The amount of energy reaching at a footpoint is not the total energy because a part of the reconnection energy will go outward and only the rest will come downward to the solar surface. Further, the energy coming toward the solar surface will be distributed into two footpoints. Therefore, the energy at a footpoint can be written, \cite{2004ApJ...611..557A} as, 

\begin{equation}
S=\frac{\epsilon}{2\pi}B_{\rm c} B_{\rm phot} v_{\rm ribb}
\label{E-poy2}
\end{equation}
  
\noindent where, factor $\epsilon$ is ($0\le \epsilon \le 0.5$). In the above equation, the coronal magnetic-field ($B_{\rm c}$) is yet to be determined. The numerical calculation of $M_i$ and $S$ requires $B_{\rm c}$. To replace $B_{\rm c}$ by the photospheric magnetic-field $B_{\rm phot}$,  \inlinecite{2004ApJ...611..557A} considered that $B_{\rm c} = a B_{\rm phot}$, where $a$ is a constant to be measured. The Poynting flux then can be written as,

\begin{equation}
S=\frac{\epsilon\,a}{2\pi}B_{\rm phot}^2 v_{\rm ribb}.
\label{E-poy3}
\end{equation}
  
\noindent From the magnetic-field extrapolation (using IDL\tm~package \rm{MAGPACK2}\footnote{\url{http://solarwww.mtk.nao.ac.jp/sakurai/en/magpack2.shtml}}, \opencite{1982SoPh...76..301S}) \inlinecite{2004ApJ...611..557A} found the ratio of coronal to photospheric magnetic-fields (\ie, $a$) $\approx$ 0.2 for the X2.3 flare event  of 10 April 2001. Using this procedure, we extrapolated the GONG LOS magnetogram (\cf, Figure~\ref{fig:neut_line}, right panel) to find the constant $a$ for the X17/4B flare event of 28 October 2003 at 11:10 UT. This turns out to be $\approx$ 0.09, which is an average value over the whole region, and is smaller in strong field area. 
  
\par To calculate the magnetic-reconnection [Equation (\ref{E-rec_rate})] and energy-release rate [Equation (\ref{E-poy3})], we require the strength of the photospheric magnetic-field. For this, we estimate the photospheric magnetic-field ($B_{\rm phot}$) from GONG LOS magnetogram at the center of the two edges of a H$\alpha$ flare-ribbon corresponding to $v_{\rm ribb}$ measurements. The estimated $B_{\rm phot}$ is shown in Figure (\ref{fig:dvmrp1}b)\spn(\ref{fig:dvmrp6}b). The reconnection parameter ($\dot{\Phi}$) and energy-release rate ($S$) are shown in Figure (\ref{fig:dvmrp1}c)\spn(\ref{fig:dvmrp6}c) with solid and dashed lines, respectively.

\inlinecite{2009SoPh..254..271X} studied flare-separation speeds and reconnection rate of the 10 April 2001 solar flare and reported a weak negative correlation between the ribbon-separation speed and the longitudinal magnetic-flux density as found here. However, contrary to the event studied by them, we do not find similarity in time profiles of the magnetic-reconnection rate and the flare-ribbon separation speed during the evolution of the X17/4B flare. In fact, we find that the magnetic-reconnection rate is better correlated with photospheric LOS magnetic-field strength.

\par The average Poynting flux over the entire RCS during the flare is found to be of the order of $10^9$. This flare lasted for over $10^4$ seconds. The approximate area of RCS estimated from EUV is $10^{19}$ cm$^2$. Therefore the energy-release is of the order of $\approx 10^{32}$ ergs, which is sufficient to produce the HXR sources (\cf, Figure~\ref{F-tr284}d).  
%
%
%
%
%
\section{Summary and Conclusions}
 \label{S-concl}
  
We present an automated technique to extract the flare-ribbons on the basis of mathematical morphology. Using the assumptions that flare-ribbons separate perpendicularly to the neutral line, we developed an automatic algorithm to compute the motion of flare-ribbons. Using flare-ribbon separation and photospheric magnetic-field, we computed the magnetic-reconnection rate and energy-release rate during the large solar flare X17/4B in NOAA 10486 observed in  H$\alpha$ at the USO.
 
\par This event occurred in a complex active region showing both extended ribbon structures as well as several fragmentary kernels. The flare was associated with filament eruption, two-ribbon separation, and a very fast CME. Temporal evolution of separation between different portions of the two main ribbons is derived. Nearly all of these show a rapid increase of velocity reaching a maximum in the range 20\spn60 km s$^{-1}$ in the initial stage, \ie, 10:58\spn11:05 UT.  Thereafter, during the next ten minutes,  the separation speed decreased at different rates depending on the magnetic-field structure and strength in the region. Interestingly, the magnetic-flux showed a rapid decrease/increase in some locations at the time of maximum velocity of ribbon separation.
\par Magnetic-reconnection in corona, as represented by the electric fields ($E_{\rm c}$) in the reconnecting current sheet, has been estimated from the measured ribbon separation speed and magnetic-fields obtained from GONG. Electric fields reaching up to 35 kV m$^{-1}$ were found in some locations. The evolution of these parameters provides evidence that the impulsive flare-energy release is indeed governed by the fast magnetic-reconnection in the corona. The magnetic-energy release rate (\ie, Poynting flux) has also been deduced for various parts of the flare. An average Poynting flux  during the flare is found to be of the order of $10^9$, which gives the energy-release of the order of $\approx 10^{32}$ ergs over the entire flare duration of over $10^4$ seconds. 
\par In summary, we have found that: \textit{i}) the entire flare-ribbon does not move with the same velocity, but different parts of flare-ribbons move with widely varying speeds in different directions, \textit{ii}) the flare-ribbon separation is observed to be decelerated in the regions of strong magnetic-fields, and \textit{iii}) magnetic-field reconnection (\ie, electric-field strength) and energy-release rate (\ie, Poynting flux) increase with increasing separation velocity.

\par A complete and accurate flare-energy release calculation is beyond the capability of the available observational and theoretical techniques. The 3D topology of active regions is only now beginning to be observed by recent projects such as the \textit{Solar TErrestrial RElations Observatory} (STEREO) spacecraft. The aim of the present study is to provide a relatively simple, automated approach to estimate the approximate level of energy-release in two-ribbon flares.

%
%
%
%
%
%
\begin{acks}

This work utilizes data obtained by the Global Oscillation Network Group (GONG) project, managed by the National Solar Observatory  which is operated by AURA Inc., under a cooperative agreement with the National Science Foundation. The GONG data were acquired by instruments operated by the Big Bear Solar Observatory, High Altitude Observatory, Learmonth Solar Observatory, Udaipur Solar Observatory, Instituto de Astrofisico de Canarias and Cerro Tololo Inter-American Observatory.

\end{acks}
%
%
%
%
%
\appendix

\textit{Coordinates of the points of a straight line:} Suppose, we have to find coordinates of the points of a straight line RS, with points R and S having coordinates ($x_1, y_1$) and ($x_2, y_2$), respectively. The distances between abscissa and ordinates of the two points will be, $ n_x=\rm abs(\textit x_2-\textit x_1) ~ \rm{and} ~ \textit n_y= abs(\textit y_2-\textit y_1)$, respectively. Let, $n=max(n_x,n_y)$. Now       
\[
\begin{array}{l}
 \rm{if}~\textit{nx}  \ge \textit{ny} ~\rm{then} \\ 
 \hspace{1cm} \rm{if}~\textit{x1}  =\textit{x2} ~\rm{then}~\textit{xi}  =\textit{x1} \,\rm{and}~\textit{yi}  =\textit{yi}  + \textit{i} \\ 
 \hspace{1cm} \rm{if}~\textit{x1}  \ne\textit{x2} ~\rm{then}~\textit{yi}  = \textit{m} (\textit{xi}  - \textit{x1})  + \textit{y1}  \\ 
 \rm{else} \\ 
 \hspace{1cm} \textit{x1}  = \textit{x2}~\rm{and}~\textit{y1}  =\textit{y2}  \\ 
 \end{array}
\]
where, $m={(y_2-y_1)/(x_2-x_1)}$ is the gradient of line RS and $ i =1,~n$ are the indices of the coordinates of points of the line RS.     
%
%
%
%
\bibliographystyle{spr-mp-sola-cnd} 
\bibliography{ms}    

\end{article}
\end{document}